# ExSched: Solving Constraint Satisfaction Problems with the Spreadsheet Paradigm

**Siddharth Chitnis, Madhu Yennamani, Gopal Gupta**
*Department of Computer Science*
*The University of Texas at Dallas*
*Richardson, TX 75080*

**Abstract.** We report on the development of a general tool called *ExSched*, implemented as a plug-in for Microsoft Excel, for solving a class of constraint satisfaction problems. The traditional spreadsheet paradigm is based on attaching arithmetic expressions to individual cells and then evaluating them. The *ExSched* interface generalizes the spreadsheet paradigm to allow finite domain constraints to be attached to the individual cells that are then solved to get a solution. This extension provides a user-friendly interface for solving constraint satisfaction problems that can be modeled as 2D tables, such as scheduling problems, timetabling problems, product configuration, etc. *ExSched* can be regarded as a spreadsheet interface to CLP(FD) that hides the syntactic and semantic complexity of CLP(FD) and enables novice users to solve many scheduling and timetabling problems interactively.

## 1 Introduction

Resource allocation has become considerably important in today's increasingly complex world. It is a key factor for success in businesses as it facilitates cost cutting and hence improves profit. These resource allocation problems are generally NP-hard problems and are solved by being modeled as constraint satisfaction problems. Thus, typical instances of a resource allocation problem---course scheduling, examination time-tabling, employee scheduling, product configuration---are all modeled as CSPs. However, most of these problems as well as other related problems such as executing business rules, product configuration, crypt-arithmetic puzzles are hard to solve using conventional means. Constraint based languages were designed to overcome these difficulties by relying on powerful propagation and optimization techniques. These techniques prune the search space and make the search for a solution faster. Consequently, languages such as CLP(FD) emerged as powerful means for solving NP-Hard Problems.

Although CLP(FD) has emerged as a powerful paradigm, many people still use the traditional paper and pencil approach to solve the scheduling and timetabling problems. One of the strong reasons for this is the syntactic and semantic complexity of constraint languages. The fact that many of the constraint satisfaction problems can be elegantly modeled as a table of constraints prompted us to develop an interface that looks like a spreadsheet. A front end with this idea was developed previously in Java which was called the Knowledgesheet. The traditional spreadsheet paradigm is based on attaching arithmetic expressions to individual cells and then evaluating them; but the Knowledgesheet interface instead allowed finite domain constraints to be attached to individual cells that are then solved to obtain a solution. This extension provides an easy-to-use interface for solving a large class of constraint satisfaction problems, i.e., those whose specification and solution conforms to a 2-dimensional structure (e.g., scheduling problems, timetabling problems, product configuration, etc.). Knowledgesheet laid foundation to an entirely different perspective of solving constraint satisfaction problems.

Microsoft Excel Spreadsheet, on the other hand, has been used for a long time for a variety of applications. It is popular because of its user-friendly interface. The popularity of Microsoft Excel and its similarity to Knowledgesheet prompted us to link Microsoft Excel and CLP(FD). In this project we develop a plug-in for Microsoft Excel called "ExSched". ExSched acts as an interface between CLP(FD) and Microsoft Excel (CLP(FD) engine, in our case SICStus, being the backend and Excel being the front end.) and thus enables even novice users to solve constraint satisfaction problems interactively.

Exsched (written in VBA) has many options that facilitate user friendliness. These options shield the complex SICStus CLP(FD) syntax from the user. The tool works in four main steps:
1. Takes user input/constraints from the excel spreadsheet and parses through it.
2. Converts the input into an executable SICStus program.
3. Executes the generated SICStus program using the back end SICStus engine.
4. Displays the result of execution in the spreadsheet.

The rest of the paper is organized as follows: In Section 2 we introduce constraint logic programming and explain its importance for solving NP-Hard problems. We then talk about the Knowledgesheet paradigm which is the foundation for this paper. Section 3 describes the Plug-In in detail. The basic idea behind ExSched and its working are explained here. Next, we illustrate the application of our tool using some day to day examples (Section 4). The sole purpose of showing these examples is to demonstrate how easily the tool can be used for solving scheduling and timetabling problems. It is followed by further details explaining how ExSched is to be used. Although this tool is functional, there are many areas that have a lot of scope for improvement. These areas are described in the future work section. It has to be emphasized at this point that the plug-in is an interactive tool that the user or programmer can use without having to have an in-depth knowledge of CLP(FD). The burden of modeling the problem in an appropriate manner to obtain proper results is still borne by the user, however, the abstractions provided by the tool help considerably.

## 2 Related Work

We start this section by introducing constraint logic programming. The advantages of constraint languages over conventional programming languages are discussed briefly. It is followed by an overview of the Knowledge Sheet paradigm. Finally, we talk about Microsoft Excel and also about the reasons that motivated this work.

### 2.1 Constraint Logic Programming

Conventional programming languages such as C and Java do not provide adequate support for solving CSPs. Traditional Logic based programming languages like Prolog on the other hand are better than conventional programming languages but are inefficient as compared to constraint logic programming (CLP) languages. This is because logic based languages use the "Generate and Test" paradigm. When the generate-and-test strategy is used, constraints are applied after the values have been chosen for the unknown variables. To ensure that the search space is pruned, the programmer has to be considerably savvy in programming the generator. For efficiency, the generator has to generate choices incrementally and in the order in which the tests can be performed.

On the contrary, Constraint languages have emerged as powerful tools to solve NP-Hard problems by using the "Test and Generate" paradigm. This strategy involves actively pruning the search space by repeating the following two steps until the solution is found [1]:
1. Propagate the constraints as much as possible
2. Choose values for unknown variables

The constraints are applied in each cycle and so superfluous searches are skipped. As a result of this, the search space and the time of execution are both reduced.

### 2.2 Knowledgesheet Paradigm

When we try to schedule events/resources the first thing that occurs to us is to model the problem into a tabular form and start solving it by a trial and error approach. Many of the scheduling and timetabling problems can be conveniently modeled as a 2-d structure. The Knowledge sheet paradigm for building programs was inspired by the spreadsheet paradigm [1] because it facilitates tabular way of representing problems.

Nevertheless, the spreadsheet by itself is not enough to solve many applications because it is limited to functions only. The main idea behind the development of Knowledgesheet was to generalize the functions to constraints/predicates [1]. And so, the spreadsheet interface coupled with the capability of solving constraints led to the creation of the Knowledgesheet.

Knowledgesheet consists of a two dimensional array of cells. Each cell acts as an unbound variable. When a problem is to be solved, it is modeled into a two dimensional structure and constraints are imposed on the cells. The knowledge sheet then gathers the input from the cells and converts it into an executable constraint logic program. The generated CLP program is then executed to print the results onto an output file. Finally, the Knowledgesheet reads the output generated and displays it in the corresponding cells.

Apart from this basic functionality, the Knowledgesheet also has some extra features which help to veil the syntactic details of CLP. Thus, the Knowledgesheet interface enables non-experts to solve CLP problems in an interactive manner.

However, there are a few disadvantages of using Knowledgesheet:
1. The user has to install the whole application (Which is an extra overhead)
2. It will take some time for the user to get accustomed to the new interface
3. Since it is a stand alone application, developing a complete version of it will be costly/time-consuming (Spreadsheet has to be implemented again)
4. The tool has to be advertised a lot to educate people (In other words it will take comparatively more time for people to know about the existence of such tools).

Instead, if Microsoft Excel is upgraded to solve constraint based problems, it will be advantageous to both excel and to the user. Excel will gain more recognition because of its increased capabilities, and it will be easier for the user to understand and use the tool as many people are already accustomed to using Excel. Also, the development time is much less because the spreadsheet interface need not be implemented again. Hence linking MS Excel and SICStus' CLP(FD) is the main objective behind the development of the ExSched system.

While ExSched is based on the Knowledgesheet system, it adds a large number of enhancements. These enhancements were inspired by our attempt to model and solve large CSPs using our "table of constraints" approach. These enhancements are described in the rest of the paper.

## 3 The ExSched Plug-In

Knowledgesheet is an interactive and convenient tool for solving many of the scheduling and timetabling problems. Nonetheless, there are some inherent drawbacks of Knowledgesheet as explained in the previous section. The main objective of this Plug-In is to overcome those pitfalls and to make the tool as simple, cost effective and user-friendly as possible.

In this section, we discuss in detail about ExSched and its working. This section is divided into three parts: (1) Visual interface. (2) Working of the tool. (3) Solving large problems interactively.

**3.1 Visual Interface**

The visual interface of this tool is the same as that of the normal Microsoft Excel spreadsheet except that it has a new menu item called "Constraint Solve" with various options. These options are divided into five major categories:
1. Macros to facilitate copying cells/formulas with and without formatting
2. Macros for some In-Built Functions
3. Macros for Arithmetic Functions
4. Macros for creating Auxiliary Tables
5. Macros for executing the problem and for viewing the solution

**Macros for Copying**

The primary function of these macros is to improve the user-friendliness. Using these macros the user can build even very complex problems just with a few mouse clicks.

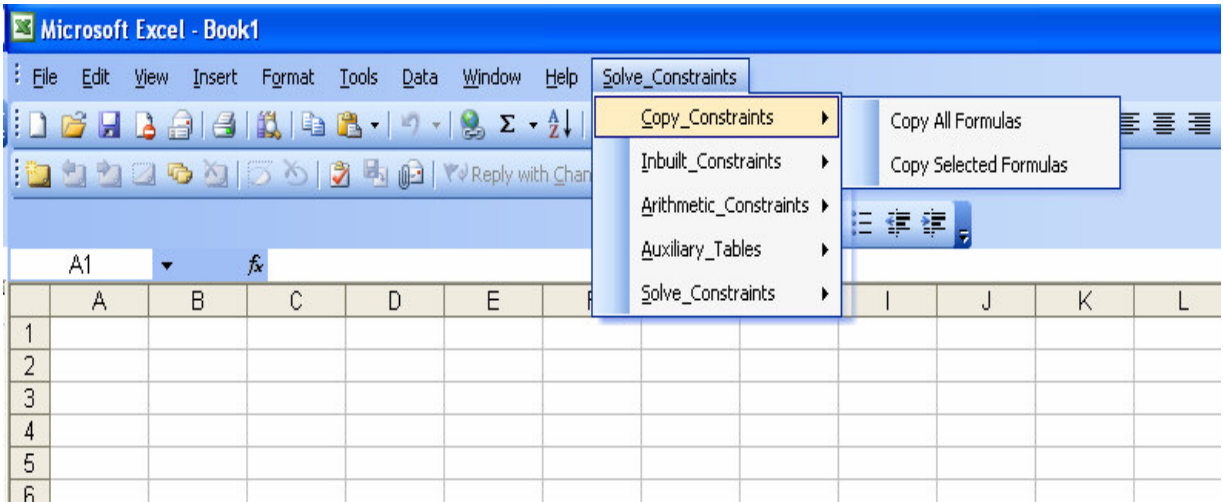

**Fig.1.** Macros for copying formulas/cells

In other words, the user rarely needs to use the keyboard. Repetitive computations are performed by copying constraints from one cell to another with appropriate transformation applied (just like how current Excel copies arithmetic expressions).There are two different ways of copying the cells as shown in Fig.1.
1. To copy all the formulae of a cell into the selected cells (appending at the end)
2. To append a formula into the selected cells

**Macros for In-built Functions**

These macros facilitate the user for framing complex constraints without having to know the syntax details of SICStus. There are four In-Built functions
1. Alldifferent-To frame a constraint that says that all the selected cells are different
2. Member-To say that a selected value should occur at least once in the selected cells
3. Frequency/Count-To restrict the number of times a value can occur in the selected cells
4. Sublist-To specify a list of values which should occur in the selected range of cells
5. IF-THEN function – To specify constraint like IF (cell1 = value1) THEN (cell2 = value2)

Fig.2. shows these four options.

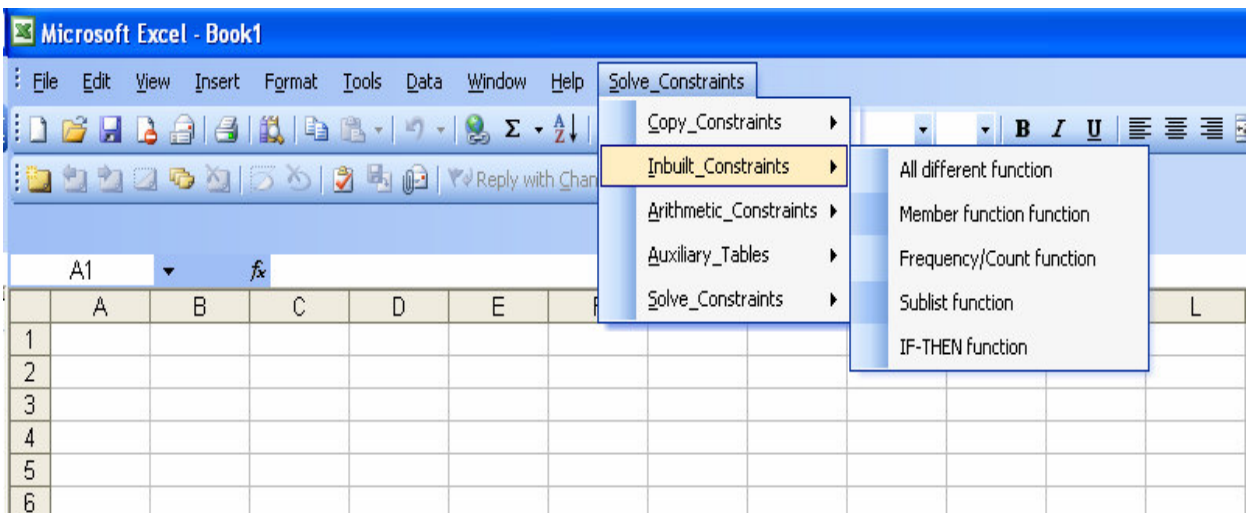

**Fig.2.** Macros for In-Built Functions

**Macros for Arithmetic Functions**

These macros assist the user for framing sum constraints without having to type them literally. The user can simply select a group of cells and use this macro to specify a constraint on the sum of the cells. There are two arithmetic functions:
1. Sum equal to a particular value.
2. Sum not equal to a particular value.

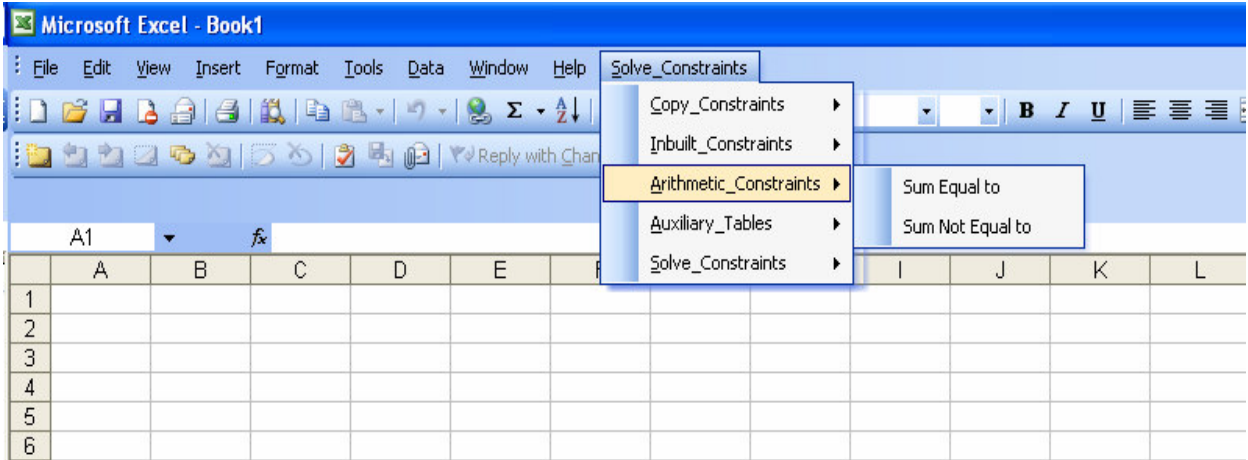

**Fig.3.** Macros for Arithmetic Functions

**Macros for Creating Auxiliary Table**

The Auxiliary tables make modeling of the problem easier and more user-friendly. These macros provide facility to create two types of Auxiliary tables.

*Map Table*

These macros are mainly used to create tables which map non-integer domain values to integer. The domain values for any variable in CLP(FD) program should be integers. But in real life scheduling problems, the domain of variables are non-integers. For example, in Employee Shift Scheduling problem, the employee can work Morning, Afternoon or Evening shift. Thus the domain of the employee variable is [Morning, Afternoon, Evening]. But in order to solve a CLP(FD) program these non-integer domain values should be mapped to some integer values say, [1,2,3]. Using Auxiliary table the user can specify this mapping once and then can use the non-integer values while actually specifying the constraints. The mapping is done automatically for those constraints using the Auxiliary table.

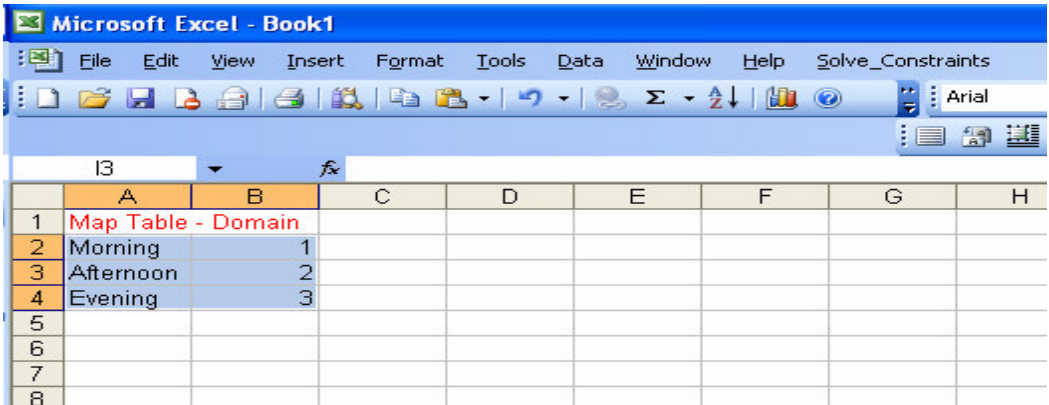

**Fig.4.** Auxiliary Table for Domain Mapping

*Fact Table*

The Auxiliary table can also be used to specify facts. The table name acts as the Predicate name and the column entries act as the arguments of the predicate.

**Macros for Solving the Constraints**

These macros are mainly used to generate the SICStus code and to execute it to see the results. There are a few extra macros like "Switch Solution" and "Load Sample Problem". The "Switch Solution" macro makes the tool interactive by enabling the user to be able to toggle back to the original constraints. In the sense, the user can try a set of constraints, see the solution and switch back to the constraints if he wants to make changes. The macro "Load Sample Problem" can be used to see some sample problems. The List of options in this sub section is:
1. Find Solution-To generate and execute the SICStus program.
2. Display Solution-To see a solution.
3. View All Solutions-To scroll through the solutions.
4. Switch Solution-To display the original constraints attached with the cells.
5. Load Sample Problem-To load sample problems.

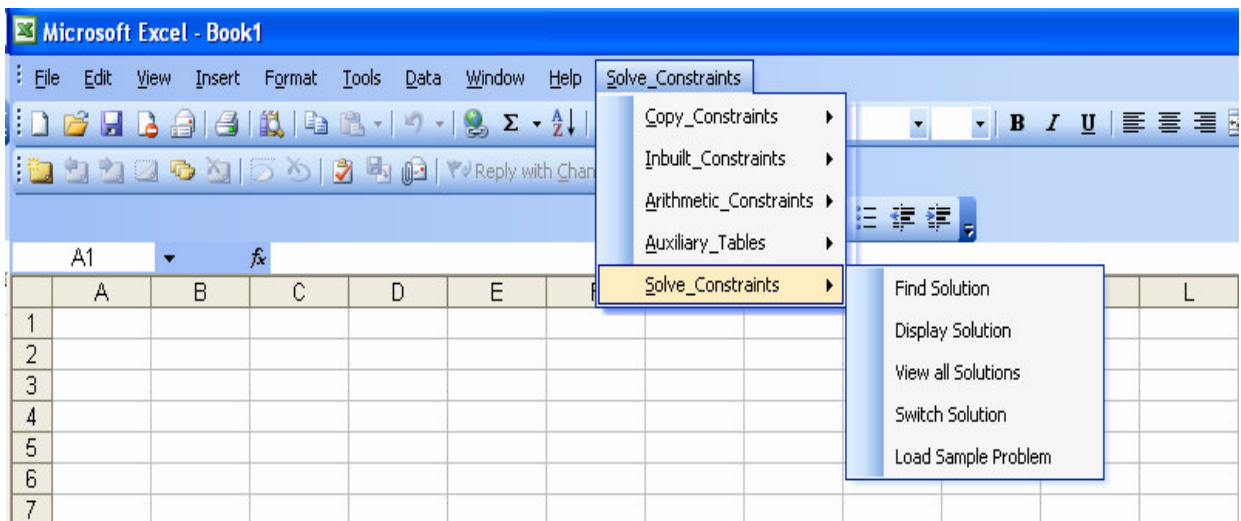

**Fig.5.** Macros for solving constraints and viewing solution

**3.2 Working of the Plug-In**

The Plug-In is basically a group of macros written in Visual Basic which works in the following way:
1. The user can enter constraints into the spreadsheet either manually or using the macros provided (The Copy macros, Arithmetic macros and the In-Built-Functions macros). When the user uses the macros, the Plug-In takes the user input and converts it into appropriate constraints. The copy macros take an additional step of formatting the formula correctly depending on the cells to which the formula is copied (making appropriate transformations).

2. When the constraints are entered and the user presses "Find Solution," the Plug-In parses through the constraints and turns it into executable SICStus program. The Plug-In transforms the constraints specified by the user into appropriate and efficient CLP(FD) constraints with correct syntax. This generation of syntactically complex CLP(FD) program is transparent to the user.
For example, the user constraint 'IF (cell1<value1) THEN (cell2 = value2)' is transformed into efficient CLP(FD) reified constraint 'cell1#<value1 #=> cell2#=value2'. This type of efficient constraint generation avoids any disjunction being entered in the constraint store. If there is a disjunction present in the constraint store then it inserts a number of choice points in the search tree. This results in huge search tree even for a reasonable size problem. This increases the time taken to search the solution in the search tree. The user doesn't have to worry about all these technical details as the efficient constraint generation is automatic.

3. The Plug-In then executes the SICStus program using a shell command. Depending on the constraints given, the SICStus program could execute immediately or take time to complete.

4. When the user presses the "Display Solution" option, the Plug-In shows the result if the result is already generated. If the result is not generated yet or if the constraints don't have a feasible solution, the "Display Solution" option prompts user to wait or relax some constraints. This macro reads the output file generated by the SICStus program and ports the values to the corresponding cells. In case the user wants to scroll through all possible solutions, the "View All Solutions" option can be used that keeps switching between all the solutions generated by the SICStus program.

5. If for some reason, the user decides to review/change/relax the constraints he has given after seeing the solution. The "Switch Solution" option can be used that reads a temporary constraints-file and writes the constraints back to the cells.

### 3. 3 Solving Large Problems Interactively

ExSched is a man-machine interface to solve constraint satisfaction problems. Both the user and the tool interact and co-operate with each other to find a solution. This interactive nature is the most important feature of this Plug-In which helps to solve large problems. Consider course scheduling at a Computer Science department that has more than 120 courses, more than 50 instructors and more than 10 classrooms. For such a huge department, the whole schedule can not be generated in one shot. However, the interactive nature of the ExSched allows users to obtain the schedule piecemeal while manually adjusting the choices. Or the user can set the instructors first, then the timings and then the classrooms. Another approach can be to divide the problem into N independent problems, solve each piece individually and then enforce the global consistency.

The tool is also very flexible. In case there is no solution for the given set of constraints (clashing constraints) or the system is taking a very long time, then the tool allows users to relax some of the constraints or reduce the size of the problem until a solution is found. The user can then gradually increase the size of the problem.

## 4 Examples

This section illustrates the ease of using this tool by explaining three examples.
1. Employee Scheduling
2. Teaching Assistant Scheduling
3. Crypt-arithmetic Puzzle

### 4.1 Employee Scheduling

Employee scheduling problem involves planning the schedule of five managers Bill, Mary, John, Gary, Linda for a week [1]. Problem description:
1. There are three shifts, morning, midday and evening, each 8.5 hrs long.
2. Each manager is supposed to work for 40 hours in a week.
3. At least one manager must be present at any point of time, morning and evening shifts together cover the whole day.
4. Each manager should get two days off.
5. Each manager works no more than 8.5 hours a day (only one shift).
6. The manager who works in the evening shift cannot be scheduled for the following day's morning shift.
7. The schedule should be fair to all managers (every manager should have at least one morning, evening and midday shift.

Fig.6. shows the constraints and Fig.7. shows the solution.

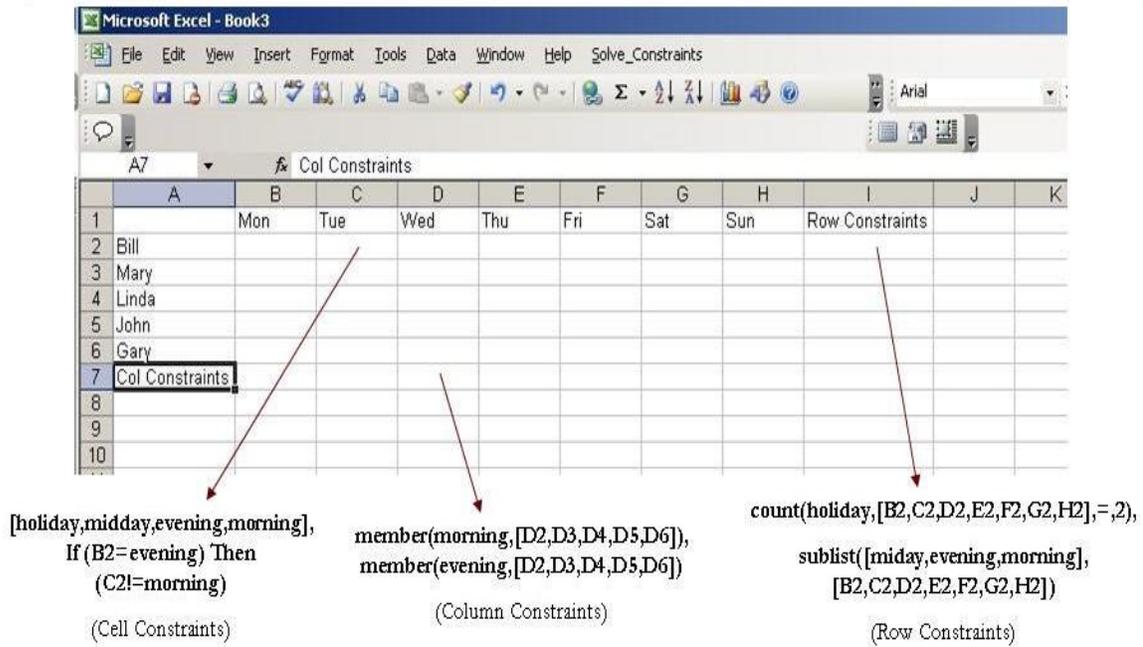

**Fig.6.** Employee Scheduling Problem with constraints

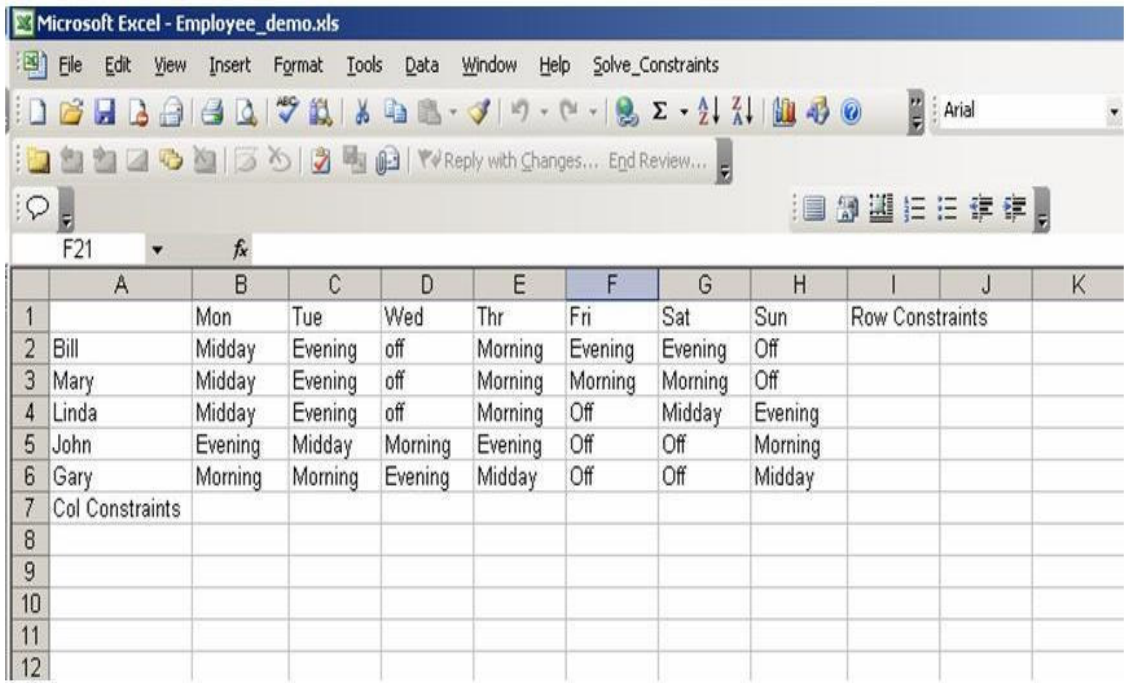

**Fig.7.** Solution for Employee Scheduling Problem

## 4.2 Teaching Assistant (TA) Scheduling Problem

TA assignment problem involves assigning TA's to courses. Problem description:
1. Each TA is assigned to a course based on the skill
2. A course will not have more than 0.25, 0.5 or 1.0 TA based on enrollment
3. A TA should not be TAing for more than 1.0
4. A TA should not be TAing more than 3 classes (so can't be broken up to 4classes for 5 hr/wk each).

Fig.8. shows the constraints and Fig.9 shows the solution.

**Fig.8.** TA scheduling problem with constraints

**Fig.9.** Solution for TA scheduling Problem

**4.3 Crypt-arithmetic Puzzle (SEND + MORE = MONEY)**

This problem involves assigning numerical values between 0 and 9 to the alphabets S,E,N,D,M,O,R,E in such a way that summation constraint (SEND + MORE = MONEY) holds. Fig.10. shows the constraints and Fig.11 shows the solution.

**Fig.10.** Crypt-arithmetic Puzzle with constraints

**Fig.11.** Solution for Crypt-arithmetic puzzle

## 5 Conclusion and Future Work

ExSched is a convenient, user-friendly and cost-effective tool to solve constraint satisfaction problems. It uses spreadsheet (MS Excel) as front end and Constraint based language (CLP(FD)) engine as back end. It provides users flexibility and interactivity to model/develop large, complex scheduling and timetabling problems. The user can solve a scheduling problem without having to know the complex syntax of the underlying Constraint Language engine. The tool exploits powerful features of the current spreadsheet such as entering formulae interactively and duplicating formulae with appropriate transformations. The tool uses these features to allow users to attach constraints to the cells. Due to this easy-to-use and popular interface non-experts can use this tool (note that managers are essentially resource allocators). ExSched is a man-machine interface: users and the tool co-operate to produce solutions. Users can give partial solutions and the rest is computed by the tool. There are many areas where there is a lot scope for improvement. The future work for this project involves:

1. Functionality to prioritize constraints - In case of large problems having a number of constraints, it is possible that there is no solution that satisfies all the constraints. Some of those constraints might be very important constraints and others might be just preferences. Thus if the user can specify which constraints are important constraints and which are not, then the tool can adjust accordingly to that and try to give solution satisfying all the important constraints and as many of the non-important constraints as possible.
2. Support for automatic constraint relaxation – in case of clashing constraints the tool can drop some of the constraints to break the tie. It can make use of the priorities of the constraints to decide which constraints to drop.